\documentclass{epl}

\title{Inter-Intra Molecular Dynamics as an Iterated Function System}

 \author{Kunihiko Kaneko}
\institute{
 Department of Basic Science, University of Tokyo and \\
 ERATO Complex Systems Biology, JST\\
   Komaba, Meguro-ku, Tokyo 153-8902, Japan \\
 }

\begin{document}

\maketitle

\begin{abstract}
The dynamics of units (molecules) with slowly relaxing internal states
is studied as an iterated function system (IFS) for the situation common
in e.g. biological systems where these units are subjected to frequent
collisional interactions. It is found that an increase in the collision
frequency leads to successive discrete states that can be analyzed as
partial steps to form a Cantor set. By considering the interactions among the units, a
self-consistent IFS is derived, which leads to the formation and
stabilization of multiple such discrete states. The relevance of the
results to dynamical multiple states in biomolecules in crowded
conditions is discussed.
\end{abstract}

Biological units (biomolecules) have internal dynamics which are often
in the same order of magnitude as those caused by interactions between
different units. Furthermore, the internal relaxation
times can be very long as was e.g. shown in recent experiments on
complex biomolecules like Ribozymes, some proteins, and so
forth\cite{Ras,others}. Besides this slowness in relaxation, the
molecules also take multiple conformations dynamically.

On the other hand, molecules in a cell are in a very crowded situation,
as for example was beautifully illustrated by Goodsell\cite{Goodsell},
leading to relatively frequent collisional interactions. Then, in contrast to the
standard situation, the time scale of the
internal relaxation is in the same order of magnitude or even slower
than that of the interaction as has also been discussed by Mikhailov and
Hess\cite{Mikhailov}.  In this case,
if the magnitude of the forces for the two are of the same order,
the interaction cannot be treated as perturbation, and such ``inter-intra dynamics" may lead to novel
behaviors unexpected from internal dynamics.
Here we investigate the possibility that
molecules with slow relaxation time scales, when put in crowded
conditions, may exhibit novel dynamical states that are not expected when
solely considering single molecular dynamics (see also
\cite{Mikhailov2}.)  To be specific, we consider the situation that the internal dynamics
have just single stable state, and seek for the possibility that just random
collisional interaction  with other molecules induce multiple stable states.

Instead of studying a realistic molecular process, we choose a rather abstract model,
to propose a novel 'crowdedness-effect' and to describe it in 
simple dynamical systems.  We will show that iterated function systems (IFS)\cite{IFS},
originally studied in the mathematics of fractals, will be
relevant to study such inter-intra dynamics, by making several extensions.

Here, we explain what we mean by ``collision" (or collisional interaction)
in the present Letter.  A biomolecule often has some binding sites.
When a specific molecule binds to such a site, conformational change in the
molecule often occurs, before it is released.  We call such a process a 'collision'
in this Letter, 
i.e., the process of binding of a specific molecule leading to
a specific conformational change and its release (see Fig.1a schematically).
By this collision, the molecules are excited so that in this crowded condition with frequent
collision, they are assumed to be in a nonequilibrium state.

Concretely, we consider situations where the average time between
molecular collisions $t_c$ is less than or in the same order
of magnitude as the internal relaxation time of each molecule $t_r$.
Instead of attempting to simulate realistic dynamics of complex
biomolecules, we consider an abstract model here in order to obtain
insights into possible novel features in systems with slow internal
relaxation and frequent interaction, as a first step to understanding
interacting biomolecule dynamics.  We take a simple internal state
of a ``molecule" $X_i$ that is represented by a
scalar variable $x_i$. For example, one can consider a toy molecule with
an internal angular variable $x$, as schematically shown in Fig. 1a. As a
starting point, we take the simplest form of relaxation of this
internal state towards $x=0$,
\begin{equation}
 dx_i/dt=-\gamma x_i +\sqrt{\gamma T}\eta (t),
\end{equation}
\noindent
with white Gaussian noise $\eta(t)$, $\gamma=1/t_r$,  and $T$ the temperature.
Hence the distribution of $x_i$ approaches Gaussian around $x =0$ as the collision frequency decreases.
In the toy molecule of Fig.1a the energy is at a minimum when the molecule is straight.

\begin{figure}[tbhp]
\begin{minipage}{.47\textwidth}
\includegraphics[width=\textwidth]{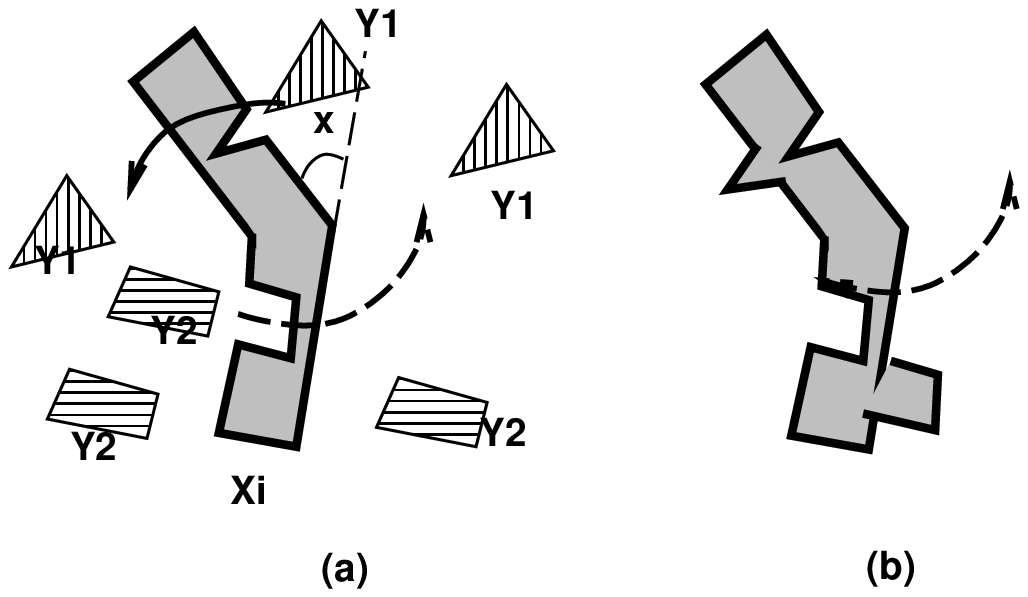}
\caption{A toy molecule, that may illustrate the internal variable $x$.  It has two binding sites,
and when the ``triangle" molecule $Y_1$ collides (binds) to the corresponding site,
a conformational change to a positive angle direction occurs, while the
``square" molecule leads to a different conformational change.
In (b), multiple units of a single type of molecule collide with each other.  When $x<0$, the ``triangle'' part
is at the outside so that it can bind to the triangle hole of another molecule more easily, and vice versa.
}
\label{fig:schem}
\end{minipage}
\hspace{0.07\textwidth}
\begin{minipage}{.4\textwidth}
\includegraphics[width=\textwidth]{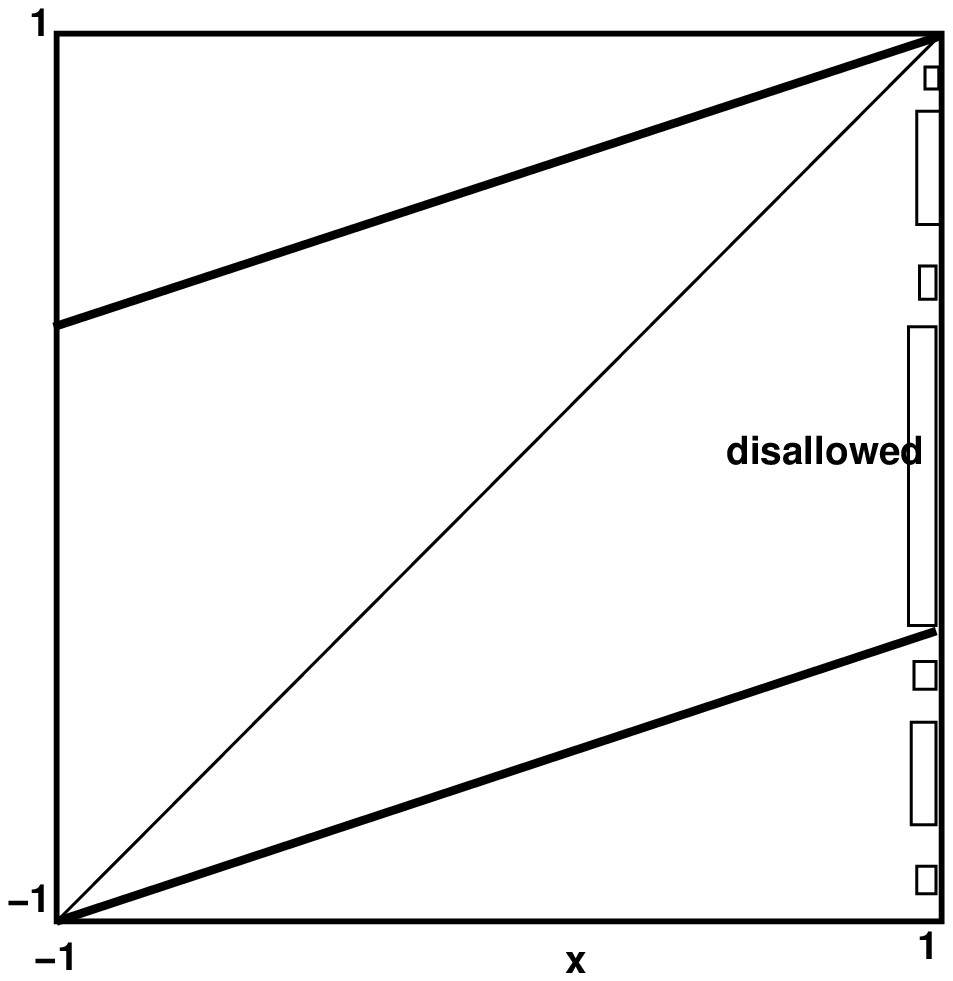}
\caption{One-dimensional map of eq.(2), with $a_1=a_2=1/3$.  
The values of $x$ which belong to the invariant
measure of the IFS consisting of these two maps can simply be obtained
by successively removing
the preimages $f^{-j}(X_k)$, starting from the interval
$X_0=[2a_2-1,1-2a_1]$, which leads to a standard Cantor set,
that is constructed by removing the middle one-third segments successively.}
\label{fig:1map}
\end{minipage}
\end{figure}


In general, collisions with other molecules induce changes in the
internal states such that $x_i \rightarrow x_{i}'$ which are assumed to
occur randomly with the rate $r_c=1/t_c$.  
As a simple idealization, let us represent this change as the map $x \rightarrow x'=f(x)$ which of
course will depend on the species of the colliding molecules. 
Representation of the change of internal state
by collision as a mapping is a simplification. Still, by writing
the collision and release of some molecule by mechanics, the internal state
after this process is given as a mapping from the initial state.  As an
abstract example, let us consider the case where there are other
molecule species $Y_1,Y_2,,,Y_k$ ( $k>1$), and by a collision, the state
of molecule $X_i$ changes as $x_i\rightarrow f_j(x_i)$ ($j=1,\cdots ,k$)\cite{note}.
Thus the collision process is represented by a mapping, whose form
depends on the molecules species that collide.  
Now we study an ensemble of molecules $X_i$, to see how the distribution of the states $x_i$ depends
on $r = r_c/ \gamma $.

Note that in the limit of $\gamma \rightarrow 0$ (i.e.,$r\rightarrow
\infty$), this problem is reduced to an IFS.  Iterated function
systems have extensively been studied in
fractal geometry, dynamical systems theory as well as in the context of image data
compression\cite{IFS,IFS3}.  It was e.g. shown that the 
stationary distribution of the states $x$ obtained through random iterations of
functions with these contracting mappings 
can take infinitely many peaks on a Cantor set.  On the other hand, in the limit of $r \rightarrow  0$
the distribution of internal state $x$ approaches just normal distribution, given by
the Langevin equation (1).  For a finite value of
$r$, and under finite temperature $T$, it is then expected that the complete 
Cantor set by the original IFS is destroyed.  Still, several states may remain,
which will be interesting since the molecule in a less crowded condition (with small $r$) takes
only a single state. If so, the formation of these multiple states are a salient feature of
``crowdedness".
Hence, we first discuss,
how the behavior in the original IFS is altered for a finite value of
$r$ and finite temperature $T$.


As a simple example, we consider the case $k=2$ with
\begin{equation}
    f_1(x)=a_1(x-1)+1; f_2(x)=a_2(x+1)-1.
\end{equation}
\noindent as illustrated in Fig.2 where
the state $x_i$ of the molecule $X_i$ moves towards
$+1$ or $-1$, depending on which molecule type $Y_j$ ($j=1,2$) it collides with.
The molecule state $x$ is shifted either to $\pm 1$ keeping some memory of the
original state value $x$ before the collision, to the degree of $a_j$ ($j=1,2$).  Thus $x$
takes the value between $x$ and $\pm 1$ with the weight $a_j:1$.
This is an interpretation of the model equation (2).
For example, in the toy model described in Fig.1, there are two types of colliding molecules:
triangles and squares.  Depending on the type, the binding site on $X_i$ where
a molecule attaches is different, so that the direction of the angle change is opposite.

Let us consider the distribution of the state values $x$ when the
collision process is repeated.  In the limit of $r \rightarrow \infty$
the distribution of the state values $x$ is shown to form a Cantor set
if $a_1+a_2<1$, i.e., if there is a gap between the maps $f_1$ and
$f_2$, as displayed in Fig.2.  (This is easily understood by considering
the distribution function invariant in time, as constructed by the
inverse of the map (2)).
This invariant measure on the Cantor set does not rely on the specific linear
form of eq.(2), but is general as long as there are (at least) two stable fixed points
around which the map is contracting (sufficiently, corresponding to the condition
$\sum_j a_j <1$).

 \begin{figure}[tbhp]
 \centering
 \includegraphics[width=.49\textwidth]{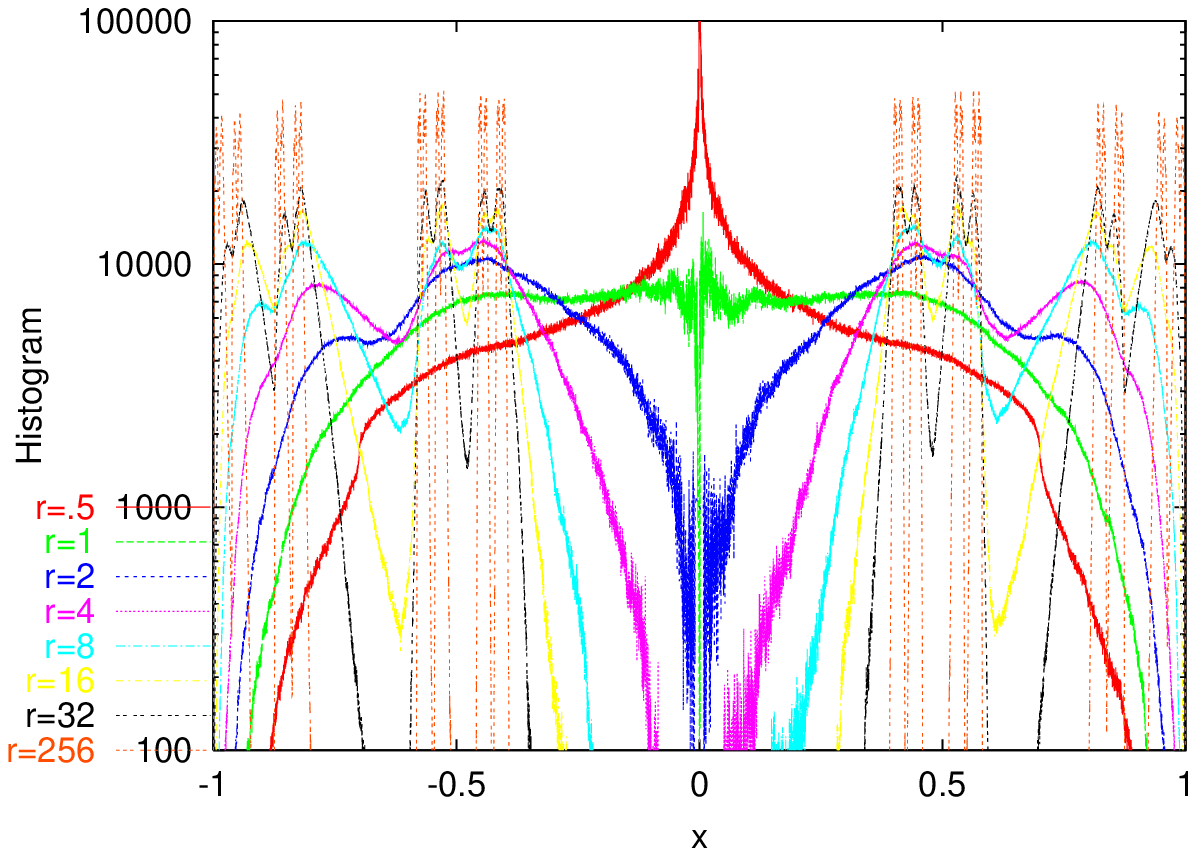}
 \includegraphics[width=.49\textwidth]{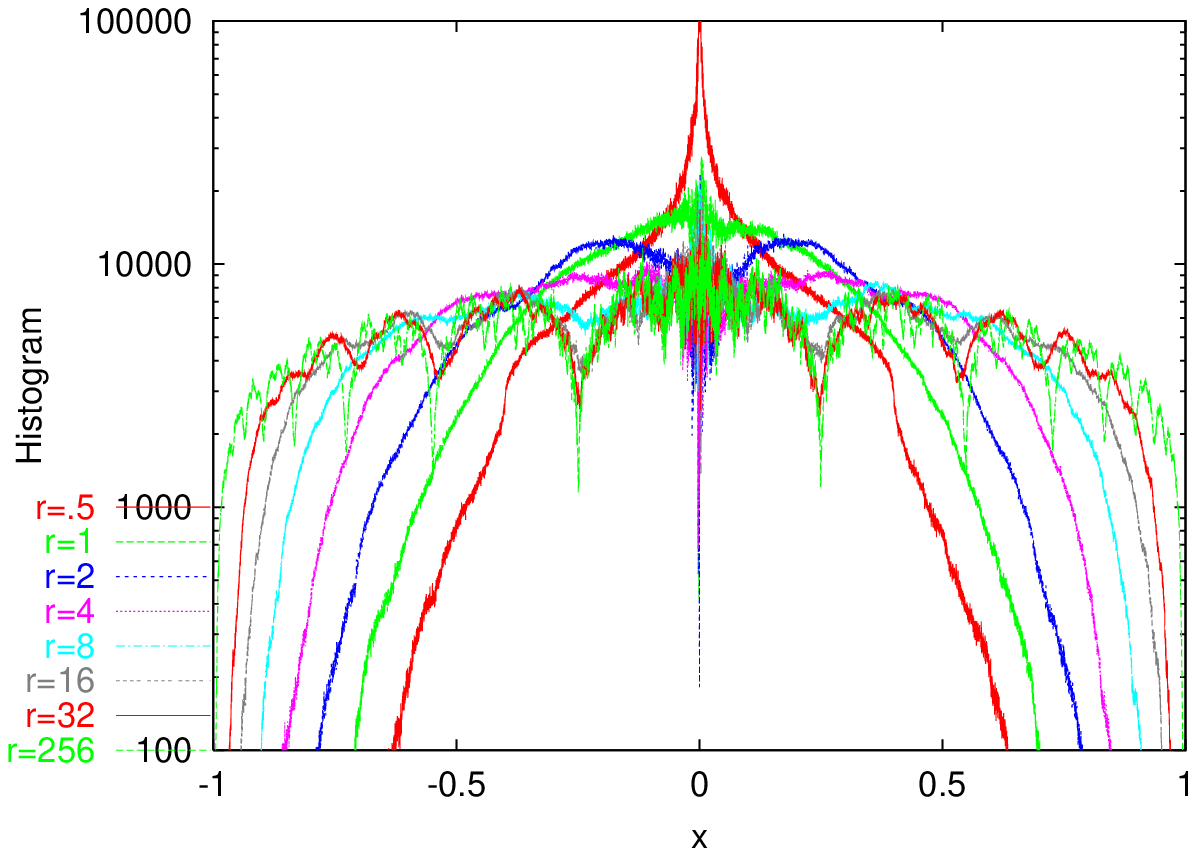}
 \caption{Histogram P(x) for various values of $r$ and fixed $\epsilon=0.01,\gamma=0.01$.
 (a)$a_1=a_2=.3$, (b) $a_1=a_2=.6$.  The histogram is obtained from
 $5\times 10^8$ iteration steps after discarding the initial transients,
 and sampled with a bin size 0.001.  $T=0.01$. }
 \label{fig:hist3}
 \end{figure}

For finite $r$ and finite temperature $T$,
we numerically studied our model by colliding two types of
equally distributed molecules with a rate $r_c$.  We obtain the
stationary distribution of $x$ by taking a large number of molecules,
or by sampling the values of $x$ of a single molecule over time.  (The results of the
two methods agree as expected from ergodicity).  As shown in Fig.3a,
several peaks are observed in the distribution with the increase of
$r>1$.  Even though the whole Cantor set structure is not observed for
finite $r$, multiple peaks are clearly discernible for $r>1$.  Several discrete
states of $x$ are formed through the molecular collisions, even if the
original relaxation dynamics has just a single stable state.  The peaks
successively split as $r$ increases, so that many discrete states are
formed.  Fig.4 shows that the number of peaks in the distribution versus
$r$ displays a power-law increase with an exponent that is consistent
with that for the increase of the peaks against the precision in the
Cantor-set construction, $-\log 2/ \log a$ (for the present example with
$a_1=a_2=a$).  In other words, an increase of the collision frequency
corresponds to an increase of the precision in the Cantor set
construction.  This power law is not altered much by changing $\gamma$
or $T$, while for large $T$ (i.e., larger noise), the increase is
suppressed.  Also, even if the condition for the Cantor set $\sum a_j
<1$ is not satisfied, the multiple broad peaks appear, as shown in Fig.3b.

\begin{figure}[tbhp]
\centering
\includegraphics[width=.47\textwidth]{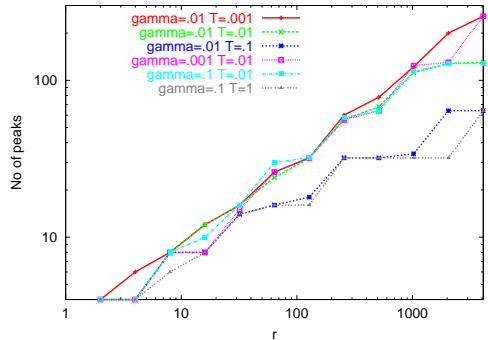}
\caption{The number of peaks in $P(x)$ plotted as a function of $r$.
$a_1=a_2=.3$.  The distribution $P(x)$ is obtained in the same way as in Fig.3.
Results from several values of $\gamma$ and $T$ varied over $0.001$, $0.01$ and $0.1$,
as displayed in the figure, are overlaid.}
\label{fig:scale}
\end{figure}

So far we have discussed the case where molecule $X$ collides with only
a finite number of other molecule species, $Y_1,..,Y_k$.  Now, we consider
the case where there is a single type of colliding molecule, but instead it
has a continuous state so that the change of molecule $X$ by it depends
on the state of the $Y$
molecule whose internal state is represented by the variable $y$.  (Or
one can consider the case that the change of $x$ by collision depends on
how it collides, e.g., the angle of the collision, which corresponds to
$y$.)  In other words, the state change of $x_i$ is given by a family of function
$g(x_i,y)$ parametrized by continuous $y$, instead of $f_j(x)$, for discrete $j$.
Here the distribution of $y$ is given
by some distribution function $\rho(y)$.  The question now is
whether the discretization of states appears even for this
continuous case. (As a dynamical system, this provides
a novel class of problem, i.e., a continuous IFS). As a specific example,
we investigate
\begin{equation}
 g(x,y)=a(x-f(y))+f(y); f(y)=tanh(\beta y)
\end{equation}
\noindent
where, for simplicity, we take $\rho(y)=constant$ over $[-1,1]$.
In the limit of $\beta \rightarrow \infty$, the model reduces to eq. (2), and here we
are interested in how the behavior is altered for finite $\beta$. 
The distribution of $P(x)$ obtained in this case is depicted in Fig.5(a)
which shows the existence of a Cantor-set type structure as $\beta $
increases beyond 2.  To have these multiple peaks in the distribution,
threshold-type dynamics ($tanh(\beta y)$ with $\beta >1$) is necessary, a situation
which often exists in molecular interactions or in biological systems in
general.

\begin{figure}[tbhp]
\centering
\includegraphics[width=.49\textwidth]{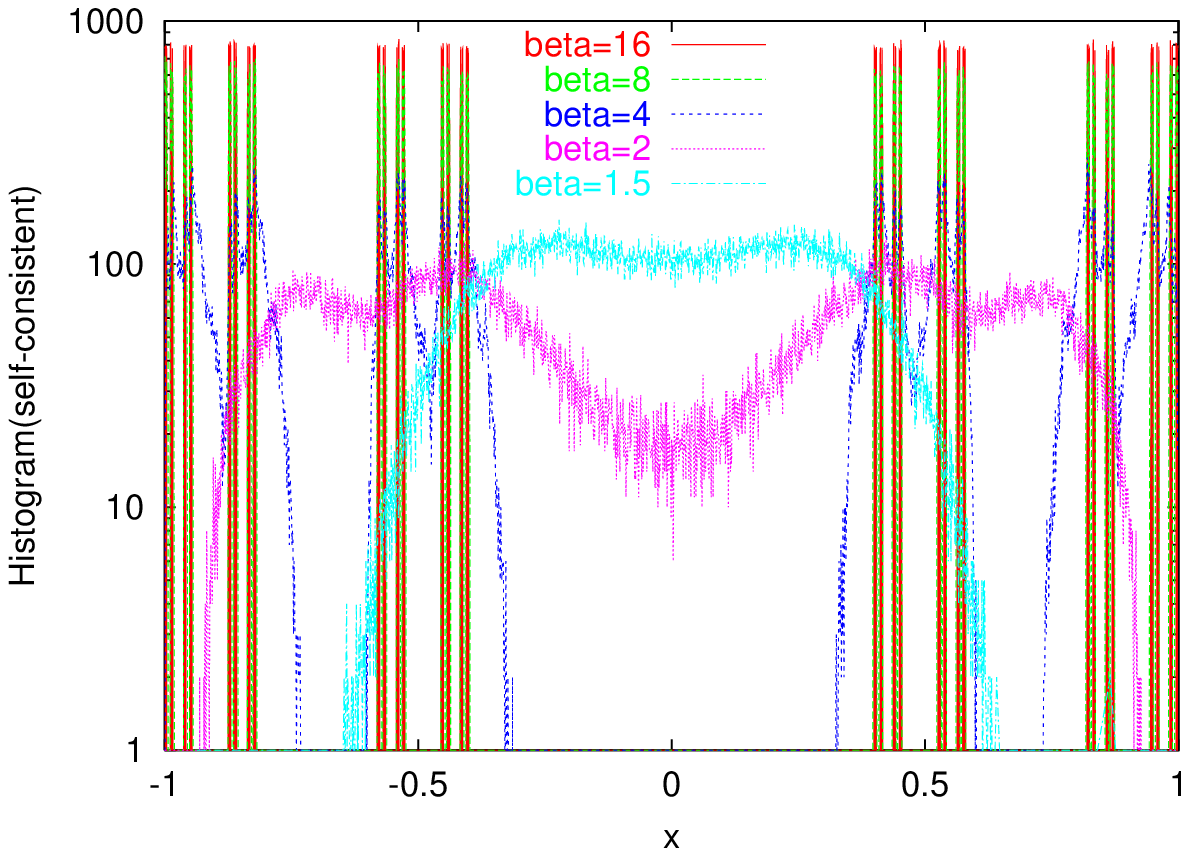}
\includegraphics[width=.49\textwidth]{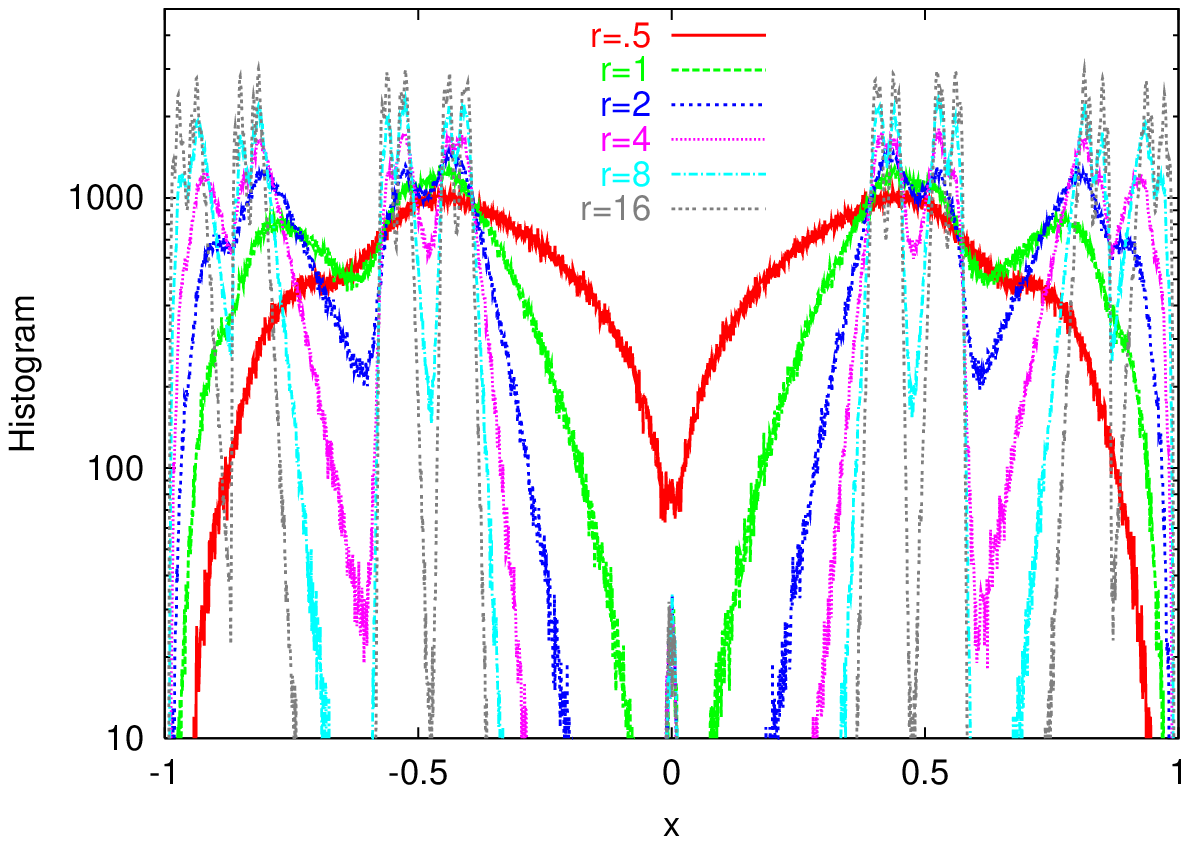}
\caption{(a)Distribution $P(x)$ of the states x, for the model (3) with $a=1/3$.
For simplicity,  the internal relaxation dynamics is not included here (only for this figure)
by taking $\gamma=0$, while, these discrete states are again stable
as long as $r_c/\gamma>1$, as discussed for Fig.3.
(b)Self-consistent formation of many discrete states.
A snapshot distribution $P(x)$ of the states $x$ is plotted
for the model (4) by using $10^6$ molecules with  $a=0.3$ and $\beta=8$. $T=0.01$, and $\gamma=0.01$.}
\label{fig:cont}
\end{figure}

Thus far we have discussed the situation where the distribution of
$\rho(y)$ is given in advance.  There are cases, however, in which the
distribution of the state $x$ influences the distribution of the state
of the colliding molecules.  For example, assume that $X$ is an enzyme
protein with multiple catalytic activities and that it depends on the
shape of $X$ which molecule it can catalyze.  Then, the production rate
of molecule species $Y_j$ depends on the distribution of $P(x)$, with
the internal variable $x$ as an index of the shape.  If this influence
of the distribution of $x$ on the distribution of $Y$ molecules is
sufficiently fast, the fraction of $Y$ molecules can be regarded to
change instantaneously and the distribution of $y$ is replaced by the
distribution of $x$.

Alternatively, one can simply consider the collision dynamics just among
$X$ molecules such that it depends on the configuration of the colliding
molecule what type of collision takes place (for example, consider the
modified toy model in Fig. 1b).  In these cases, the state change of a
molecule $X_i$ by a collision with a molecule $X_j$ is given by the
mapping $x_i\rightarrow g(x_i,x_j)$.  If the distribution of $x_j$ were
given and fixed, this would be nothing but the IFS discussed above. As
the distribution of $X$ changes by the collisions however, the problem is
represented by a "self-consistent IFS".


 As a specific example, consider the case
\begin{equation}
  g(x_i,x_j)=a(x_i-y)+y; y=tanh(-\beta x_j).  \end{equation}
\noindent The result of a numerical simulation of this model is given in
Fig.5(b) where it can be seen that the distribution has again multiple
peaks when $r$ is larger than 1, i.e. when the collisions are frequent.
With the increase of $r$, peaks successively split thus discretizing the
states. There are already 4 peaks for $\beta=2$, and as the threshold
function is steeper, more peaks are formed,  again mirroring a few
steps in the construction of a Cantor set.  These discrete states
corresponding to peaks are stabilized ``self-consistently"
through interactions with other $X$ molecules and
are stable  against noise and the influence of the relaxation dynamics.

To sum up, we have shown that multiple discrete states can be formed as
partial Cantor sets, through collisions of molecules, even if the single
molecule dynamics has just a single stable state.  This process is
possible when the time scale of the collisional interaction is similar to or faster
than the internal relaxation time scale and when there are several types of
interactions that cause different conformational changes.  In this sense, observed is
a novel class of phenomena, i.e., 
the formation of multiple internal states due to the crowdedness of molecules.
The discretization of states becomes clearer as the ratio of collision frequency
to internal relaxation time increases.  Note that although this Cantor
set structure is not complete, several steps of this structure are
observed as discrete states, despite the presence of internal relaxation
and noise.  Hence it is demonstrated that iterated function systems,
originally introduced as abstract mathematical models, are relevant to
molecular dynamics, while an extension to a continuous family of
functions in the IFS is also introduced.
Furthermore, when there is a feedback to the distribution of the
internal states from the distribution of the colliding molecules, a
novel class of statistical dynamics is introduced that could be termed a
self-consistent IFS (SIFS).  Note that in the limit of $r \rightarrow
\infty$,
the SIFS is nothing but a random-update version of a globally coupled map\cite{GCM},
where the distribution of the states $x_i$ can show collective motion.
By taking, for example, a non-monotone map of $g(x_i,x_j)$ for $x_i$, there
are cases that the distribution function changes in time.

Another extension of the present approach will be an explicit use of the
population dynamics of the colliding molecules $Y_j$, instead of the
adiabatic elimination of $Y_j$ molecules as adopted in the SIFS.  For
example, by assuming that the type of molecule $Y_j$ is synthesized
depending on the state $x$, the production rate of the molecules $Y_j$
is proportional to $\int_{x \epsilon I_j} \rho(x,t)dx $, where
$\rho(x,t)$ is the density of the state $x$ at a given time $t$, and
$I_j$ is the range of $x$ values that catalyzes the synthesis of $Y_j$.
By introducing the population dynamics
 \begin{math}
 dN_j/dt= c\int _{x \in I_j}\rho(x) dx -\Gamma  N_j
 \end{math}
for molecule species $Y_j$, and the collision dynamics $x_i
\rightarrow f_j(x_i)$ as in eq.(2),  
self consistent dynamics on the
distribution $\rho(x,t)$ is derived that couples with the population
dynamics of $N_j$ as above.  This model was also studied numerically and
it is found that, besides the appearance of multiple peaks in the
distribution, the height of the peaks can change over time, suggesting
the existence of the collective motion well-known to occur in globally
coupled maps.

 As for the use of the relaxation dynamics, one might question the
origin of the relaxation itself since one can argue that it occurs
through collisions with other molecules such as water molecules.  In
general, the time scale or the size of such 'heat-bath' particles
causing relaxation should be much smaller than that of the colliding
molecules $Y$, and if this is indeed so, then the model investigated
here will be justified.  

On the other hand, it would also be interesting
to consider a model in which the relaxation process is treated as
resulting from collisions with other particles representing the
heat-bath (e.g., water molecules).   Numerical
study of such model again showed the emergence of discrete peaks
validating the overall approach advocated here.

The aim of the present Letter is the proposition of a theoretical
framework where (self-consistent) IFS are applied to molecular dynamics.
As an illustration a very simple model for collision and relaxation
dynamics is discussed here, while a variety of straightforward
extensions, such as higher-degrees of freedom for internal dynamics,
more realistic collision dynamics and feedbacks with several types of
molecules, will be discussed in the future.  Indeed, the kinds of
molecular dynamics occurring in cells are much more complex since there
is a huge variety of colliding molecule types and high-dimensional
relaxation dynamics even with regards to single biomolecules.  It will
be interesting to investigate how the proposed concept is relevant to,
or is extended to, such cases.
On the other hand, it is interesting to note that
multiple conformations in Ras proteins have been observed, and that these
are thought to have different functions in a cell\cite{Ras}.  Although further
studies are necessary, the mechanism studied here suggests that the
conformations of such plastic proteins change depending
on the concentrations of the molecules themselves and on the molecules
they collide with.  We expect to see the stabilization of the discrete
states through the interaction with other molecules, as well as the
switching among these states.

Finally, the present discretization of states by IFS can also be
applied to systems with internal dynamics and 'kicking' interactions in
general\cite{Broomhead}.  The dynamics of coupled neurons can also be
discussed along these lines, while the relevance of Cantor sets in
neural information processing has been pointed out by Tsuda\cite{Tsuda}.
Several extensions of IFS as proposed here will provide novel
theoretical schemes for analyzing biological systems that exhibit the formation of
stable, discrete states through synergies between interaction and internal dynamics.

 The author would like to thank H. Takagi, T. Yanagida, Y. Sako,
Y. Sato, S.Sasa and F.H. Willeboordse for illuminating discussions.


\begin{thebibliography}{999}
\bibitem{note}
Of course, fluctuations can be added as a noise term but here we have
not done so as the effect of fluctuations can be considered incorporated
in the term $\eta(t)$, and consequently no qualitative change is
introduced by adding noise to $f_j(x_i)$.  Also, finite time interval for collision, i.e.,
between binding and release, can be introduced, which again does not
lead to essential change of the result.
\end{thebibliography}
 \end{document}